# The Associated Volume sampling algorithm as an alternative method for the calculation of ionisation cluster size distributions in computational nanodosimetry

**João F. Canhoto[1,2,*], Yann Perrot[3], Reinhard Schulte[4], Ana Belchior[2,5], Carmen Villagrasa[3]**

1. Departamento de Física, Instituto Superior Técnico, Universidade de Lisboa, Av. Rovisco Pais, 1049-001 Lisboa, Portugal

2. Centro de Ciências e Tecnologias Nucleares ($C^2TN$), Instituto Superior Técnico, Universidade de Lisboa, Estrada Nacional 10, 2695-066 Bobadela LRS, Portugal

3. Autorité de sûreté nucléaire et de radioprotection (ASNR), PSE-SANTE/SDOS/LDRI, F-92260, Fontenay aux Roses, France

4. Division of Biomedical Engineering Sciences, Loma Linda University, Loma Linda, CA 92350, USA

5. Departamento de Engenharia e Ciências Nucleares, Instituto Superior Técnico, Universidade de Lisboa, Av. Rovisco Pais, 1049-001 Lisboa, Portugal

* Corresponding author: joaofcanhoto@tecnico.ulisboa.pt

## Abstract

In computational nanodosimetry, Monte Carlo Track Structure (MCTS) simulations are employed to calculate ionisation cluster size distributions (ICSDs), which are crucial for characterising mixed radiation fields at the nanoscale. The Uniform Sampling (US) algorithm, commonly used for this purpose, is inefficient when evaluating ICSDs conditioned on clusters exceeding a given size $\nu$ threshold. This study investigates a more efficient alternative—the Associated Volume (AV) algorithm—against the US approach for computing conditional ICSDs ($\nu \geq 1$) from proton tracks simulated with Geant4-DNA. Two configurations of the AV algorithm were evaluated: the standard AV–Overlap, allowing sensitive volumes to overlap (with a $1/\nu$ correction), and the novel AV–No Overlap, which prevents overlap. We compared the conditional ICSDs, mean cluster size ($M_1$), complementary cumulative frequencies ($F_k$ for $k \in [2,7]$), and overall execution time per run and per event. Deviations exceeding two standard deviations of the mean difference were considered statistically significant. Although statistically significant differences in ICSDs, $M_1$ and $F_k$ were observed, the relative differences in the AV–Overlap rarely exceeded 5%, whereas the No Overlap configuration they ranged from <1% to approximately 15% at the higher $F_k$. Both AV configurations achieved execution times one to two orders of magnitude shorter than the US algorithm. Our findings indicate that the AV–Overlap algorithm may be a preferred alternative for calculating conditional ICSDs and derived nanodosimetric quantities, offering substantial gain in computational efficiency without compromising accuracy.

**Keywords:** Nanodosimetry, Monte Carlo track-structure simulation, Geant4-DNA, computational algorithms

# 1. Introduction

Monte Carlo (MC) simulations are numerical computational techniques that employ statistical random resampling and are particularly valuable for solving, among others, problems that would be either analytically intractable or experimentally impossible, time-consuming, or too costly (Harrison 2010; Naqa et al. 2012). In radiation physics, MC Track Structure (MCTS) codes, such as Geant4-DNA (Incerti et al. 2010a, 2010b, 2018; Bernal et al. 2015), Topas-nBio (Schuemann et al. 2019) and PHITS track-structure (TS) mode (Sato et al. 2024), have become established tools in micro- and nanodosimetry to study the interaction of ionising radiation with micro- and nanoscopic targets, respectively, such as a cell nucleus or the DNA molecule (Incerti et al. 2016). Differently from the general-purpose codes, like PENELOPE (Baró et al. 1995), EGSnrc (Kawrakow et al. 2000) or Geant4 (Agostinelli et al. 2003; Allison et al. 2006, 2016), which use the Condensed History (CH) approach (Dingfelder 2012), TS codes simulate the physical interactions of charged particles, particularly electrons, often produced as secondary particles, on an interaction-by-interaction basis and down to the eV or sub-eV thresholds (Nikjoo et al. 2006). The result is a detailed description of the particle track, including the spatial distribution of physical interaction points and their nature (Dingfelder 2012).

Nanodosimetry focuses on the characterisation of the ionisation component of the particle track structure with nanometric resolution, either by simulations (MCTS codes) or experimental measurements with nanodosimeters, and seeks to establish the metrological basis for characterising biological and biomedical effects of radiation fields (Rabus et al. 2014; Palmans et al. 2015; Conte et al. 2018). The fundamental quantity in nanodosimetry is the ionisation cluster size, $\nu$, a stochastic quantity that represents the number of ionisations produced by a primary particle and its secondaries in a specific nanometre-sized target volume. The particle tracks that compose a radiation field are characterised using the frequency distribution of $\nu$, also called the ionisation cluster size distribution (ICSD, $f(\nu)$), and associated nanodosimetric parameters (Palmans et al. 2015).

In computational nanodosimetry, different algorithms have been used to calculate ICSDs, and we can generally group them into two classes: "sampling" (e.g., Uniform Sampling (Ramos-Méndez et al. 2018)) and "clustering" (e.g., DBSCAN algorithm (Francis et al. 2011)). In this work, we will focus on the former. Sampling algorithms for nanodosimetry use sensitive volumes (SVs), usually representing small DNA volumes, to score the number of ionisations occurring near the particle track, within a defined scoring region.

The most common sampling algorithm in nanodosimetry is the Uniform Sampling (US). In this algorithm, the centre position — and, if needed, the orientation — of the sampling volumes is determined using a uniform random number generator. These positions can be generated either before or after the simulation, but the key point is that they are independent of the location of interaction points (e.g., ionisations). The works of Alexander, Bueno and Ramos-Méndez (Bueno et al. 2015; Alexander et al. 2015; Ramos-Méndez et al. 2018) are notable examples of the application of US in nanodosimetry, where non-overlapping cylinders were used to calculate nanodosimetric ICSDs. This algorithm, however, is relatively inefficient and time-consuming. This is particularly evident when calculating conditional ICSDs ($\nu \geq 1$), i.e. distributions under the condition that all volumes scored at least one ionisation, since most sampling volumes will not score any ionisation.

For this reason, this work turns attention to the Associated Volume (AV) sampling algorithm. Originally proposed by Lea in his seminal book "Actions of Radiations on Living Cells" (Lea 1946), the AV algorithm has been adopted in microdosimetry since the late 1980s and presented as a more efficient method for obtaining microdosimetric spectra (Kellerer 1985; Rossi and Zaider 1996; Famulari et al. 2017). While Famulari used spherical and cylindrical sensitive volumes with sizes down to the nanometre scale, these works did not apply the AV algorithm to calculate nanodosimetric ICSDs. The current work aims to fill this gap.

The AV algorithm is based on the concept of "associated volume of the track", which is defined as the union of all spheres of radius $r$ that are centred at transfer points (e.g., ionisations). From this definition, any sphere of radius $r$ whose centre lies within the associated volume will score at least one ionisation ($\nu \geq 1$). The algorithm is an iterative procedure, where each iteration consists of 1) randomly selecting an ionisation, 2) placing the centre of a sphere of radius $r$ at a uniformly randomly generated point within a distance $r$ from the selected ionisation, and 3) counting the number of ionisations inside the sphere.

The main goal of this work is to investigate the AV algorithm and compare it with the standard US algorithm in radiobiological and biomedical applications, where the sensitive volumes, such as DNA, do not overlap. Two versions of the AV algorithm were investigated: one allowing overlap of sensitive volumes ("AV-Overlap") but correcting for overlap when it is not possible, and another that computationally excludes overlap of sensitive volumes ("AV-No Overlap"). Both versions were compared with each other and with the US algorithm, considered the gold standard of sampling algorithms, in terms of ICSD accuracy and computational efficiency for primary protons spanning a wide range of initial energies. Finally, we discuss the usefulness of these algorithms in biomedical applications along with their limitation in predicting only conditional ICSDs.

# 2. Materials and Methods

## 2.1. Algorithms and Sensitive Volume

### 2.1.1. Sensitive Volume

In nanodosimetric studies motivated by radiobiological or biomedical applications, it is common practice to approximate the sensitive volume by a cylinder with a height of 3.4 nm and a diameter of 2.3 nm, which represents a ten base pair DNA segment (Grosswendt 2005; Lazarakis et al. 2012). In this work, however, we chose a sphere as the biologically relevant sensitive volume for computational reasons. Unlike cylinders, spheres have no preferential axis, thereby eliminating the need to generate random orientations during placement. Furthermore, as the height of the reference cylinder (3.4 nm) corresponds to the accepted critical distance between two strand breaks on opposite DNA strands to form a double-strand break (DSB) — a form of deleterious DNA damage—, we also adopted 3.4 nm as the diameter of the spherical sensitive volume.

The following subsections provide detailed information about modifications and unique aspects we have introduced in each algorithm for our study.

### 2.1.2. Uniform Sampling

In "Uniform Sampling" (US), sampling volumes are placed and, if necessary, oriented randomly inside the scoring region, usually in a non-overlapping fashion.

In this work, $10^6$ non-overlapping spheres of 3.4 nm in diameter were randomly placed inside the scoring region (1 μm$^3$ cube). A sensitive volume was deleted and resampled if it overlapped with another. During the placement procedure, a sequential numeric ID (SVID) was assigned to each sensitive volume. The location of the sensitive volumes was the same for all events.

Whenever an ionisation occurred (by the primary or secondary particles) inside one of the sensitive volumes, we stored the current *eventID* and the volume's *SVID* in the output file.

In our work, an event corresponds to the transport of one primary particle and its secondaries until their energy falls below a threshold or the particle leaves the world volume. The *eventID* is a numerical, sequential integer number that identifies each event.

### 2.1.3. Associated Volume

In the AV algorithm, sampling volumes are placed, iteratively, in the scoring region at random locations that are no more than a distance $r$ apart from an ionisation. In our work, as indicated in the beginning of section 2.1, $r$ was set to 1.7 nm. Figure 1 displays a schematic representation of an AV iteration.

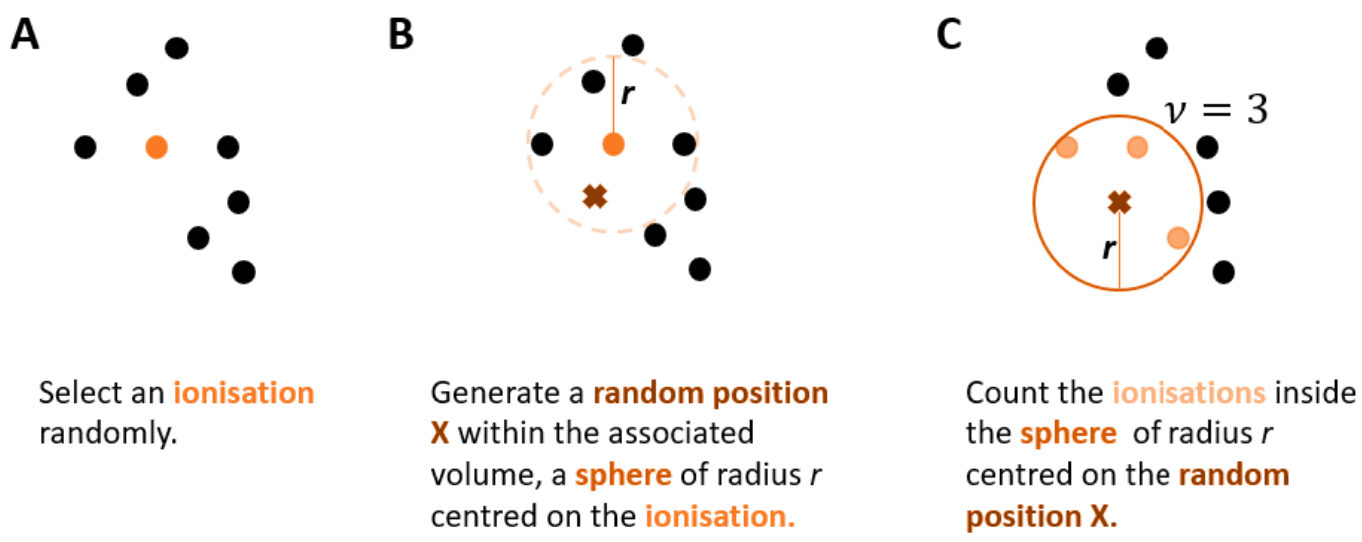

Figure 1. Steps in a single iteration of the *"*Associated Volume*"* sampling algorithm: A) an ionisation is randomly selected (orange dot) from *all* ionisations (black dots); B) A point (X) in the sphere of radius $r$ (dashed circle) centred at the selected ionisation (orange dot) is randomly generated; C) If the necessary conditions are met (depending on the chosen configuration), the centre of the sensitive volume of this iteration (a sphere of radius $r$) is placed at point X, and the ionisations inside this sensitive volume are counted*;* in the example, there are three ionisations, resulting in a cluster size $\nu = 3$.

The standard implementation of this algorithm allows sensitive volumes to overlap. However, in radiobiological and biomedical applications of nanodosimetry, overlap is not possible assuming sensitive volumes represent non-overlapping DNA segments. Inspired by previous work in microdosimetry (Kellerer 1985; Rossi and Zaider 1996; Famulari et al. 2017), we used the following weighting procedure to correct the ICSD for overlap:

$$f(\nu) = \nu^{-1} f^{*}(\nu), \tag{1}$$

where $f^{*}(\nu)$ is the ICSD scored with the algorithm before the weight is applied. In this work, the data resulting from this correction are labelled as "AV-Overlap", which is understood as the standard AV algorithm corrected for overlap.

Additionally, we investigated how prohibiting overlap during execution compares to correcting for overlap. To this end, we developed a novel implementation of the AV algorithm in which sensitive volumes are not allowed to overlap (labelled "AV-No Overlap"). After generating the centre of each new sphere, the code enters a validation step to ensure the new sensitive volume does not overlap with any existing sphere. If no overlap is detected, the sensitive volume is retained; otherwise, it is discarded, and a new iteration begins. At the end of each iteration, the algorithm removes the ionisations inside the newly generated sensitive volume from the list of ionisations for future iterations. It is important to note that, for this version, it is possible that some ionisations may not be included in any sensitive volume when the algorithm terminates. In the remainder of the text, these two versions of the AV algorithm will be referred to as "configurations".

Since the AV is an iterative procedure, it needs conditions that once met cause its termination. We implemented the following termination conditions:

i. The number of sensitive volumes equals the number of ionisations.

ii. No ionisations remain for random selection.

iii. The number of iterations reaches a preset maximum (in this work, the maximum number was three times the number of ionisations in the event).

Condition i. was implemented to avoid oversampling of tracks, condition ii. is self-explanatory and applies only to the No Overlap configuration, and condition iii. was implemented as a fail-safe termination to avoid an infinite loop.

The coordinates of energy depositions resulting in ionisations by the primary proton or secondary electrons in the scoring region were recorded in an array that served as input to the algorithm. At the end of the simulation run, an output file was saved containing the following information: *eventID*, *SVID* (Sensitive Volume ID, a sequential numerical ID assigned to each sensitive volume during runtime) and *clusterSize* (the number of ionisations contained in the sensitive volume).

## 2.2. Proton track simulations

### 2.2.1. Geant4-DNA Simulations

We simulated tracks of protons in liquid water using the Geant4-DNA toolkit based on Geant4 version 11.0. The G4EmDNAPhysics_option2 physics constructor was used for the transport and interaction of particles in liquid water with the proton and hydrogen elastic scattering processes turned off. Option2 was chosen because its dielectric model for ionisations of electrons provides a broader energy range (up to 1 MeV) than option4 (10 keV) and option6 (250 keV) and it still was the default constructor in Geant4-DNA examples. Although option4 is now the recommended constructor by the Geant4-DNA collaboration (Incerti et al. 2018), it is expected that the results presented will not be significantly affected by the choice of the physics constructor.

All simulations used monoenergetic protons as the primary particle. We chose protons for two reasons: 1) it is a common charged-particle used in nanodosimetric applications, such as radiobiological studies for charged-particle therapy (Ramos-Méndez et al. 2018) and radiation protection in space (Schulte et al. 2008), and 2) the physical models of protons in Geant4-DNA are comprehensive (i.e., they include charge-changing cross-sections). Protons were started with initial kinetic energies of 0.25, 0.5, 0.75, 1, 2.5, 5, 7.5, 10, 25, 50, 75, and 100 MeV. These energies were chosen to cover a broad range and achieve an approximately spatially invariant mean cluster size within the scoring region.

The protons were tracked in a 1 $\mu m^3$ cube (scoring region) made of liquid water. This volume was chosen because the vast majority of ionisations caused by a proton and its secondary particles are likely to be contained within a volume of this size (Braunroth et al. 2020) and for reasons of computational efficiency. The protons started 200 nm from the scoring volume and travelled along the $+x$ axis.

The number of primary particles ranged from $2 - 5 \times 10^3$ and $0.8 - 2 \times 10^5$ for the AV and US algorithms, respectively. For the AV–No Overlap and US algorithms, that number was obtained by performing successive simulations with increasing number of particles until the total number of clusters in the run was approximately equal to the total number of clusters obtained with the AV–Overlap. This ensured that all algorithms had compatible statistical uncertainties.

Only ionisations by the primary protons and the secondary electrons were scored, but the simulations still included other physical processes available in Geant4-DNA.

It is also important to note that, while we opted to use Geant4 for particle transport simulation, this should not be seen as a requirement. The use of any of these algorithms is independent of the track structure code used for particle transport.

### 2.2.2. Computational Efficiency

To evaluate the computational efficiency of each algorithm, we used the mean execution time as the metric. The execution time was obtained with built-in functions of Geant4. Specifically, we used the *GetRealElapsed()* function from the *G4Timer* class to extract the "real elapsed time" of each run. The timer started inside the *BeginOfRunAction()* function, after geometries were built, and stopped with the *EndOfRunAction()* function, after Geant4 had written and closed the output files. For both algorithms, the execution time included the track structure simulation and the clustering formalism; for the AV algorithm, this involved the iteration process of Figure 1, and for the US algorithm, it involved checking whether the ionisations occurred within a sensitive volume. We performed the simulations using eight cores on an Intel Core i7 CPU @ 3.40 GHz and a multithread build of Geant4 with eight threads.

## 2.3. Data Analysis

Simulations were performed three times (runs) with different initial seeds to assess the statistical uncertainty of the results (standard deviation). The algorithms sampled each track within the Geant4-DNA simulation. The output data files were processed using the R Statistical Software v4.2.3 (R Core Team 2023).

For each combination of algorithm, proton energy and run, we derived the conditional ICSD and calculated the following nanodosimetric parameters:

a) The first moment, or mean cluster size, $M_1$:

$$M_1 = \sum_{\nu=1}^{\infty} \nu f(\nu) \tag{2}$$

b) The complementary cumulative frequencies $F_k$, with $k \in [2,7]$:

$$F_k = \sum_{\nu=k}^{\infty} f(\nu) \tag{3}$$

When considering distributions conditioned on cluster sizes equal to or greater than $j$ ionisations, the complementary cumulative frequency $F_{k=j}$ is equal to one. Since we are considering distributions conditioned on cluster sizes equal to or greater than one ionisation, $F_1$ was not included in our analysis.

For each energy and algorithm, the parameter means were calculated for the three runs. Differences between the means larger than two standard deviations of the difference were considered statistically significant.

# 3. Results

## 3.1. Ionisation Cluster Size Distributions

Figure 2 displays the conditional ionisation cluster size distributions (ICSDs) obtained with the AV-No Overlap (red circles), the AV-Overlap (blue diamonds) and US (purple dots) algorithms for three of the twelve energies studied. Each point and its error bars represent the mean and two standard deviations of the mean, respectively. In the present work, we used the AV algorithm to determine conditional ICSDs and compared it with conditional ICSDs obtained using the US method. Future work will consider how the AV algorithm can also be used to calculate the unconditional ICSD in a given sensitive volume.

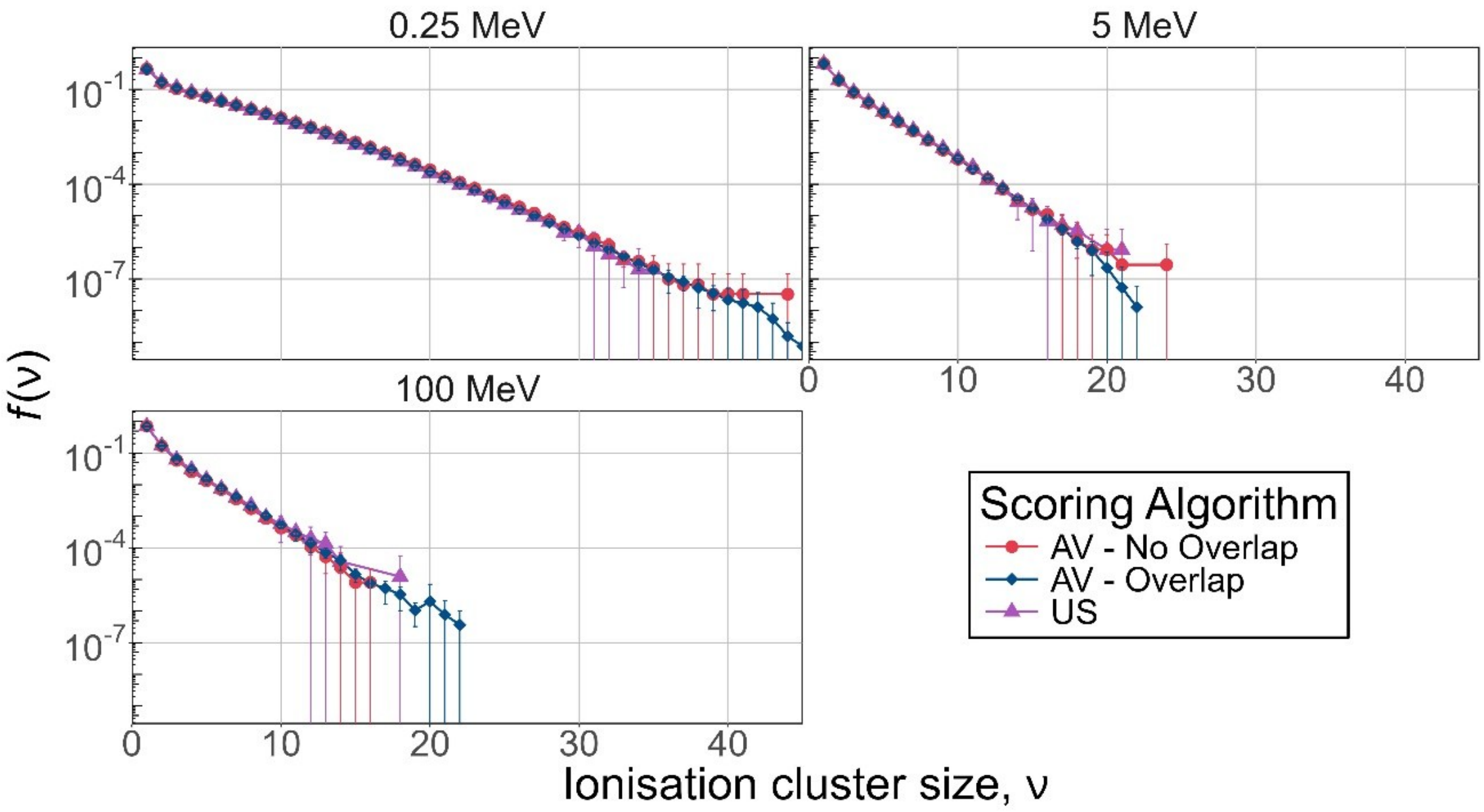

Figure 2. ICSD conditioned on cluster sizes of one or more ionisations, $f(\nu)$, for the AV - No Overlap (red circles), AV – Overlap (blue diamonds) and Uniform Sampling (purple triangles) algorithms after simulation of 0.25 MeV (top left), 5 MeV (top right) and 100 MeV (bottom left) protons. The distributions are the mean of three independent runs, and the vertical error bars indicate two standard deviations of the mean. Lines are a guide for the eye. Both axis scales are consistent across all panels.

The distributions are predominantly in good agreement within or deviated by slightly more than two standard deviations, with larger deviations observed at larger and more infrequent cluster sizes.

## 3.2. Nanodosimetric Quantities

Figure 3 shows the values of $M_1$ and $F_k$ ($k \in [2,7]$) calculated from the conditional ICSDs obtained using the AV-No Overlap (red circles), AV-Overlap (blue diamonds) and US (purple triangles) algorithms. Each point and error bar represent the mean and two standard deviations of the mean over three independent runs. Additionally, the panels in Figure 4 show the ratio of these quantities as a function of the initial kinetic energy of the primary proton, using the values of the US algorithm as the reference. The error bars represent two standard deviations of the ratio, calculated using standard uncertainty propagation.

The absolute differences in $M_1$ and $F_k$ were generally larger than two standard deviations of the mean difference. For $F_k$, these absolute differences increased with both proton energy and cluster size threshold *k*. However, as shown in the left and middle panels in the top row of Figure 4 ($M_1$ and $F_2$), the relative differences (with respect to the US algorithm) remained within $\pm 4\%$ for both AV configurations. For $F_k$, the relative differences exhibited a similar trend for both configurations, with positive differences for proton energies below 2.5 MeV, peaking at around 500 keV, and negative relative differences for proton energies above 2.5 MeV. In this region, the relative differences increased in an energy-dependent manner up to about 10 MeV, where they reached a plateau of about 10% for the AV-No Overlap configuration and 5% for AV-Overlap. This plateau was evident for $k \geq 3$ in the AV-No Overlap and for $k \geq 5$ in the AV-Overlap. Overall, the relative differences increased with $k$, ranging from less than 1% for $F_2$ to approximately 7% (AV-Overlap) and 15% (AV-No Overlap) for $F_7$. Larger statistical fluctuations were observed at higher energies due to lower event frequencies.

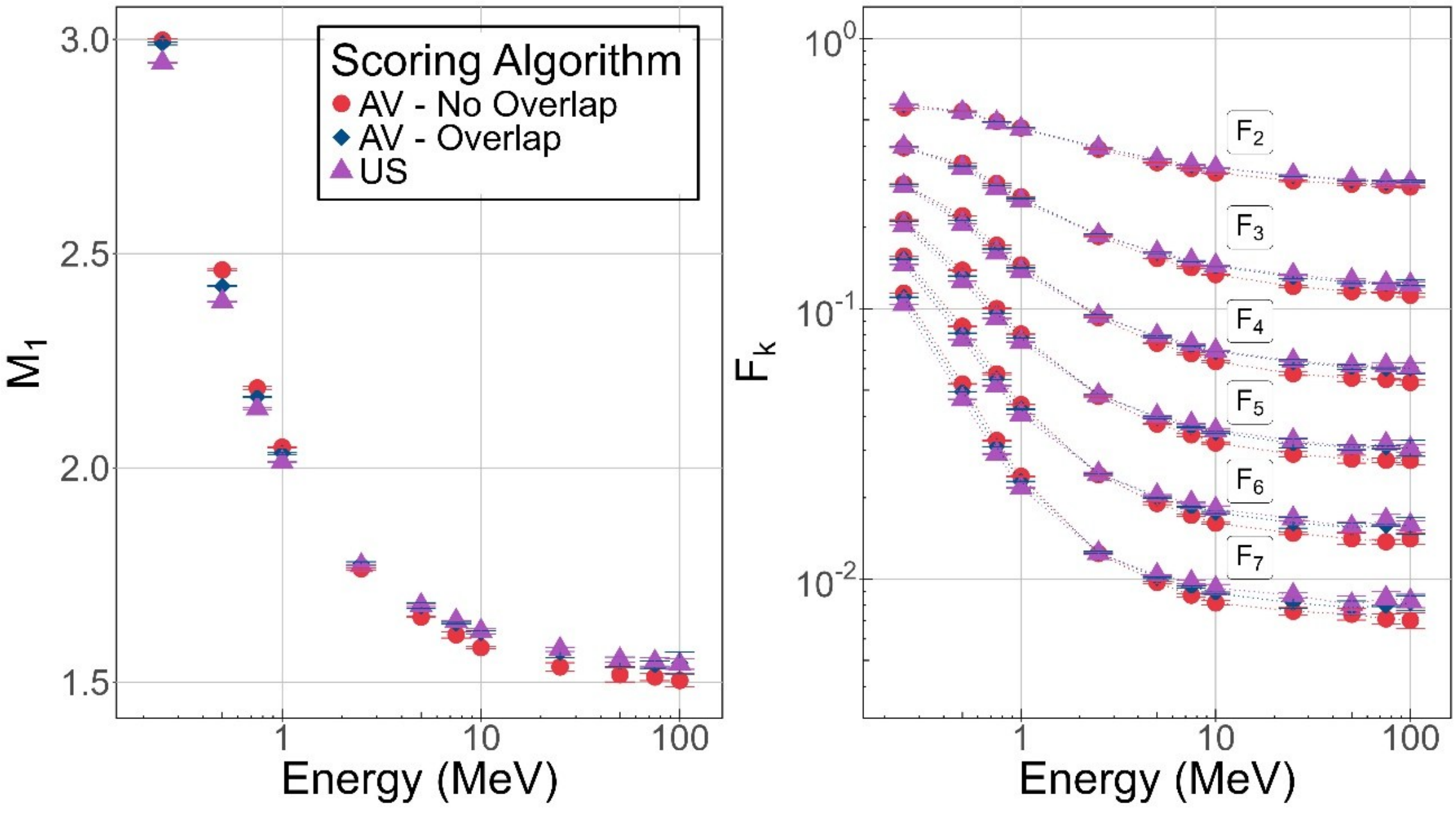

Figure 3. First moment, $M_1$, (left plot) and complementary cumulative frequencies $F_k, k \in [2,7]$ (right plot) as a function of the initial kinetic energy of the primary proton calculated from the conditional ICSDs obtained using the AV-No Overlap (red dots), the AV-Overlap (blue diamonds) and the US (purple triangles) algorithms. Lines are guide for the eye. The values are the mean of three independent runs, and vertical error bars indicate two standard deviations of the mean.

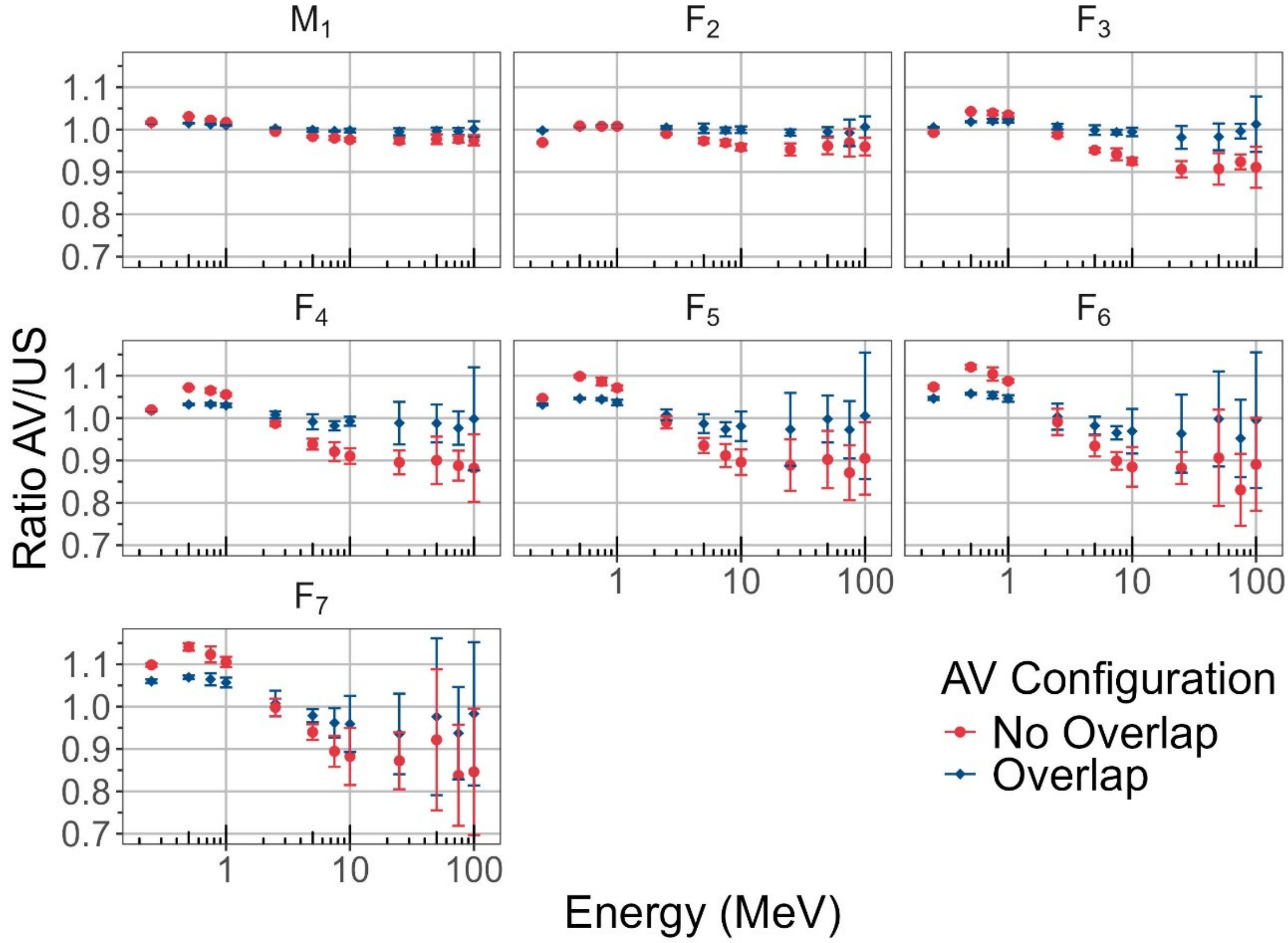

Figure 4. Ratio of the nanodosimetric quantities ($M_1$, $F_k$) for the two AV algorithms relative to the US algorithms used as reference. Red circles and blue diamonds correspond to the AV-No Overlap and AV-Overlap configurations, respectively. The panels represent $M_1$ and different $F_k, k \in [2,7]$. Note that the vertical axes show the ratios on a linear scale and the horizontal axes show the proton energies on a logarithmic scale. Vertical error bars indicate two standard deviations of the ratio.

## 3.3. Computational Efficiency

The left-hand side plot in Figure 5 presents the overall execution times of the three scoring algorithms as a function of the initial kinetic energy of the primary proton, ranging from 0.5 MeV to 100 MeV. The simulations were designed to achieve similar statistical uncertainties across all algorithms by ensuring a comparable total number of clusters in a run. For that reason, the right-hand side of Figure 5 displays the execution time per event. The values are the mean execution times, and the vertical error bars represent two standard deviations of the mean.

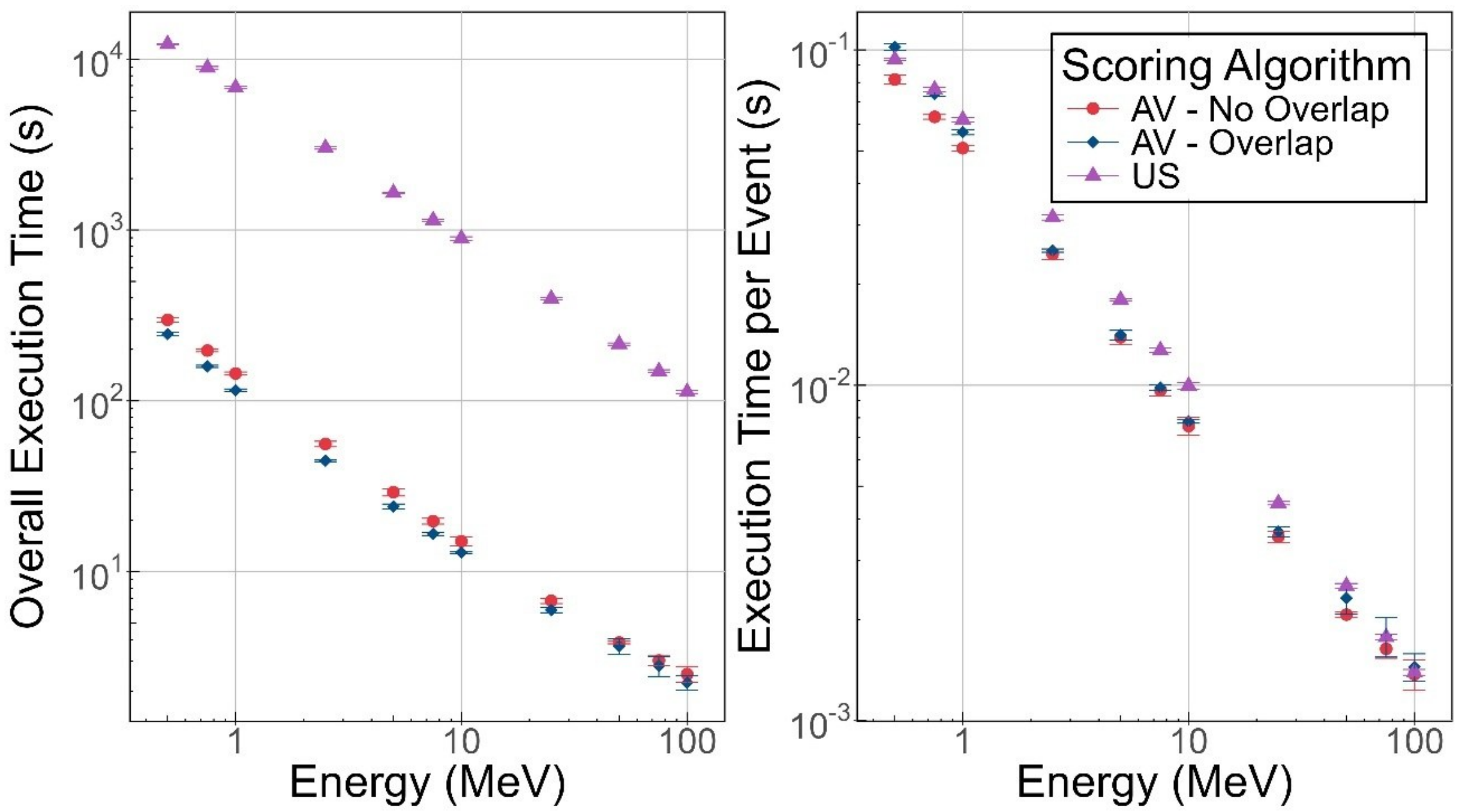

Figure 5. Overall execution time required to achieve similar statistical uncertainties (left) and execution time per event (right), in seconds, as a function of the initial kinetic energy of the primary proton, for the AV-No Overlap (red circles), AV-Overlap (blue diamonds), and the US (purple triangles) algorithms. Values are the mean of the independent runs and vertical error bars are two standard deviations of the mean.

The figure demonstrates that the US algorithm is computationally more expensive than the AV algorithm to achieve the same number of clusters in a run for similar statistical uncertainties. In this case, the AV algorithm exhibited execution times that were one to two orders of magnitude shorter, regardless of configuration. However, when considering execution time per event, a different trend emerged. For the two highest energies, differences between the three algorithms were not statistically significant. In the range of 1 MeV to 50 MeV, the US algorithm had a significantly higher execution time per event than the AV, while the two AV configurations were statistically equivalent. A distinction between the AV configurations became evident only at energies below 1 MeV, where the No Overlap configuration showed faster execution times. The Overlap configuration performed as poorly as or worse than the Uniform Sampling algorithm only for the highest LET protons. It is important to note that the plotted execution times of the US algorithm did not include the time spent placing the sensitive volumes, which can be a significant part of the total execution time.

# 4. Discussion

This study aimed to investigate two versions of the Associated Volume (AV) sampling algorithm as alternative methods for the calculation of conditional ionisation cluster size distributions (ICSDs) and compare them among themselves and with the Uniform Sampling (US) algorithm, commonly used in computational nanodosimetry, with respect to accuracy and computational efficiency. The first configuration of the AV that was investigated allows sensitive volumes to overlap (AV–Overlap) but requires a correction when applied to scenarios where overlap is not possible, such as in sensitive DNA volumes. We chose the correction factor previously used for microdosimetric spectra (Kellerer 1985; Rossi and Zaider 1996; Famulari et al. 2017) but never tested for calculating ICSDs in nanodosimetry. The second configuration tested an alternative computational approach that inherently prevents volume overlap (AV-No Overlap). We compared conditional ICSDs and derived nanodosimetric quantities and evaluated execution times as a measure for computational efficiency for all three algorithms.

The ionisation cluster size distributions obtained with the three algorithms were generally in close agreement, although in some cases, the mean absolute differences exceeded two standard deviations (Figure 2Figure 2. ICSD conditioned on cluster sizes of one or more ionisations, $f(\nu)$, for the AV - No Overlap (red circles), AV – Overlap (blue diamonds) and Uniform Sampling (purple triangles) algorithms after simulation of 0.25 MeV (top left), 5 MeV (top right) and 100 MeV (bottom left) protons. The distributions are the mean of three independent runs, and the vertical error bars indicate two standard deviations of the mean. Lines are a guide for the eye. Both axis scales are consistent across all panels.). While statistically significant, these differences were relatively small in absolute terms. Variations in cluster size distributions propagate into the derived nanodosimetric parameters, $M_1$ and $F_k$ ($k \in [2,7]$), and, again, absolute and relative differences (with respect to the US algorithm) were generally small, though they showed a systematic dependence on proton energy (Figure 3, Figure 4). Relative differences in $M_1$ between the US and the two versions of the AV algorithms were consistently smaller than those observed for $F_k$, ranging from less than 1% to about 4%. In contrast, relative differences in $F_k$ ranged from less than 1% to about 5% for the Overlap configuration and from less than 1% to about 15% for the No Overlap configuration. A clear trend of positive and negative differences was observed for energies below and above 2.5 MeV, respectively. This trend could be attributed to the bias of the AV algorithms in selecting larger over smaller clusters for the high LET region of protons and reversed bias for the low-LET protons. The bias was weaker for the Overlap configuration, suggesting that allowing for overlap and then correcting it results in smaller differences (relative to the US) compared to prohibiting overlap. The bias for large clusters is most noticeable at lower energies and larger $k$ because, under these conditions, large clusters dominate the ICSD.

One should emphasise, from a radiobiological point of view, that larger clusters are more important than small or medium-sized clusters because they are more difficult, or even impossible, to repair by DNA repair mechanisms. Although, up to which value of $k$ such clusters are important and should be considered in characterising radiation quality is currently unknown, efficient scoring of these clusters provides an important advantage. Both versions of the AV algorithm demonstrated much higher efficiency in scoring large clusters compared to the US algorithm, making them particularly valuable for radiobiological and biomedical applications where such clusters are critical.

For all energies, simulations with the US algorithm had execution times one to two orders of magnitude higher than those with the AV algorithms (Figure 5, left plot). This a consequence of its low sampling efficiency (number of volumes with at least one ionisation relative to the total number of volumes), which results in an increased number of events per run required to achieve the same number of clusters. Except for the highest and lowest energies, the US also had longer execution times per event than the AV algorithms (Figure 5, right plot). This may stem, among other factors, from how Geant4 handles particle tracking in the presence of initially placed SVs, leading to more simulated steps compared to the AV.

The AV-Overlap configuration showed slightly faster overall execution times than the AV-No Overlap configuration for energies below 50 MeV (Figure 5, left plot), but it was less efficient per event for energies below 1 MeV (Figure 5, right plot). These results suggest that the Overlap configuration is generally more CPU-demanding for low-energy (high LET) particles. The faster overall execution time of the Overlap configuration stems from its ability to generate more clusters in an event, reducing the number of simulated tracks needed to achieve a given number of clusters in a run. However, this benefit comes at the cost of slower execution time per event, particularly for high LET particles – characterised by densely packed ionisation clouds–, since the algorithm only finishes when the number of sensitive volumes equals the number of ionisations. In contrast, the No Overlap configuration benefits from additional termination conditions, allowing it to complete events more quickly. Finally, our results also suggest that the process of checking for overlap had a negligible effect on the execution time for our specific sensitive volume.

In summary, although the No Overlap configuration was designed to conceptually resemble the US algorithm, our results revealed systematic deviations dependant on proton energies, suggesting it introduces a bias for larger and lower clusters for highest and lowest LET protons, respectively. Additionally, this configuration also adds computational complexity due to the need for overlap checks, which becomes particularly challenging to calculate and implement for geometrically anisotropic shapes (e.g., cylinders). In contrast, the Overlap configuration, while allowing volume overlap, applies a well-defined correction and delivered results more consistent with the reference algorithm. Our findings confirm that the AV algorithm, especially in the Overlap configuration, offers a practical and computationally efficient alternative to US algorithm for calculating conditional ICSDs and derived nanodosimetric parameters. The comparable accuracy and significantly reduced computation time make it particularly well-suited for time-intensive track-structure simulations and for generating large libraries of ICSDs, such as those needed for the novel concept of cluster dose (Faddegon et al. 2023).

While we have shown that the AV-Overlap is an efficient alternative to the US algorithm, other studies have explored methods to make the US algorithm more computationally efficient, such as combining it with particle splitting variance reduction techniques (Ramos-Méndez et al. 2017). These techniques could potentially be integrated with the AV-Overlap algorithm as well, possibly leading to further computational gain.

In this work, we chose a spherical sensitive volume of 3.4 nm for reasons of computational ease and radiobiological significance. It is important to note, however, that our conclusions are likely dependent on the shape and size of the sensitive volume, and both should be selected based on the specific application. While this study does not directly address these dependencies, future work should investigate which shape and size of sensitive volumes best reflects the frequency of radiobiological endpoints. For example, it may be possible to retain the computational advantages of spherical volumes while selecting a size that produces results equivalent to a cylindrical SV of 2.3 nm diameter and 3.4 nm height.

Recently it was pointed out that there are two conceptually different interpretations of the ICSD, $f(\nu)$, resulting from the overlay of track structure and sensitive volumes (Braunroth et al. 2020). In the standard interpretation, $f(\nu)$ represents the probability distribution of $\nu$ in a set of disjoint sensitive volumes, while in the alternative interpretation, $f(\nu)$ represents the probability for finding an SV that receives $\nu$ ionizations when randomly placed around the track. One should note that the ICSD derived from the US algorithm corresponds to the first interpretation and the ICSD derived from the AV algorithm corresponds to the second interpretation. Our findings show that, in practice, these two interpretations yield very similar ICSD and derived nanodosimetric parameters.

It is important to note that the AV algorithms only provide *conditional* ICSDs, whereas applications of nanodosimetry in radiobiology and biomedicine for mixed radiation fields (e.g., in charged-particle radiation therapy) generally require *unconditional* ICSDs. This motivates the development of an efficient method to derive unconditional ICSDs from the AV algorithm. This work is currently in progress.

# 5. Conclusions

This study compared two versions of the Associated Volume (AV) sampling algorithm (Overlap vs. No Overlap) with the standard Uniform Sampling (US) algorithm for calculating conditional ionisation cluster size distributions (ICSDs) and derived nanodosimetric parameters for protons of varying energies. Our results show that both AV algorithms offer superior computational efficiency without compromising the accuracy in ICSDs and nanodosimetric quantities compared to the US algorithm and an advantage of the Overlap configuration in terms of both accuracy and efficiency for most proton energies. Overall, these findings highlight that the AV-Overlap algorithm is an efficient and reliable alternative to the US algorithm, particularly when computational resources are limited or when large libraries of ICSDs are needed, such as in application of nanodosimetry to charged-particle radiation therapy.

# Data availability

Data will be made available on request.

# Credit authorship contribution statement

**João F. Canhoto:** Formal analysis, Methodology, Software, Visualisation, Writing – original draft. **Yann Perrot:** Conceptualization, Formal analysis, Methodology, Writing – review & editing. **Reinhard Schulte:** Formal analysis, Supervision, Writing – review & editing. **Ana Belchior:** Resources, Writing – review & editing. **Carmen Villagrasa:** Conceptualization, Formal analysis, Methodology, Supervision, Writing – review & editing.

# Acknowledgements

João F. Canhoto acknowledges financial support from Fundação para a Ciência e a Tecnologia (FCT) through the research grant PRT/BD/151544/2021.